\documentclass[aps,preprint]{revtex4}%
\usepackage{amsfonts}
\usepackage{amsmath}
\usepackage{amssymb}
\usepackage{graphicx}%
\providecommand{\U}[1]{\protect \rule{.1in}{.1in}}
\begin{document}
\title[Competition between FFLO and BCS in an optical potential.]{Competition between the Fulde-Ferrell-Larkin-Ovchinnikov phase and the BCS
phase in the presence of an optical potential}
\author{Jeroen P. A. Devreese$^{1}$}
\author{Michiel Wouters$^{1}$}
\author{Jacques Tempere$^{1,2}$}
\affiliation{$^{1}$TQC (Theory of Quantum systems and Complex systems), Universiteit
Antwerpen, B-2020 Antwerpen, Belgium.}
\affiliation{$^{2}$Lyman Laboratory of Physics, Harvard University, Cambridge, MA 02138, USA.}

\begin{abstract}
In three dimensional Fermi gases with spin imbalance, a competition exists
between Cooper pairing with zero and with finite momentum. The latter gives
rise to the Fulde-Ferrell-Larkin-Ovchinnikov (FFLO) superfluid phase, which
only exists in a restricted area of the phase diagram as a function of
chemical potential imbalance and interaction strength. Applying an optical
potential along one direction enhances the FFLO region in this phase diagram.
In this paper, we construct the phase diagram as a function of polarization
and interaction strength in order to study the competition between the FFLO
phase and the spin balanced BCS phase. This allows to take into account the
region of phase separation, and provides a more direct connection with
experiment. Subsequently, we investigate the effects of the wavelength and the
depth of the optical potential, which is applied along one direction, on the
FFLO state. It is shown that the FFLO state can exist up to a higher level of
spin imbalance if the wavelength of the optical potential becomes smaller. Our
results give rise to an interesting effect: the maximal polarization at which
the FFLO state can exist, decreases when the interaction strength exceeds a
certain critical value. This counterintuitive phenomenon is discussed and the
connection to the optical potential is explained.

\end{abstract}
\date{\today}

\pacs{03.75.Ss, 05.70.Fh, 74.25.Dw}
\maketitle

\section{Introduction}

In the past decennium, the field of cold atoms has witnessed the experimental
realization of various new quantum coherent phenomena in ultracold gases
\cite{Review Dalibard}. Amongst the various systems that are being studied,
fermionic superfluids occupy a prominent role \cite{Greiner, Chin, Zwierlein
Ketterle}. Contrary to bosonic gases, fermionic gases must form pairs between
fermions in different spin states in order to form an s-wave superfluid. A
fundamental question that has attracted wide attention recently, is what
happens when the ratio between these different spin-states is altered, thereby
creating a spin-imbalanced or polarized superfluid. This effect is akin to
applying a magnetic field to a superconductor, which leads to electrons
aligning their spins. In superconductors, and in condensed matter systems in
general, it is difficult to control various parameters such as the
polarization or the interaction strength. In ultracold gases, however, one has
an unprecedented degree of control over these parameters \cite{Bloch review,
Greiner Q and A}. For instance, experimentalists can tune the interaction
strength by making use of Feshbach resonances (for a review on this subject
see Ref. \cite{Chin Grimm}). Using this technique, the crossover from a
Bardeen-Cooper-Schrieffer (BCS) superfluid to a Bose-Einstein condensate (BEC)
of bound molecules has been studied \cite{Regal Greiner, Bourdel Salomon,
Partridge Hulet, De Melo}. Furthermore, polarization can be created by first
loading particles into one hyperfine state and subsequently applying a
radio-frequency sweep to send a controlled number of particles to a different
hyperfine state. This experimental liberty has led to the realization of many
experiments on the imbalanced ultracold Fermi gas \cite{Zwierlein Ketterle 2,
Shin Ketterle, Partridge Hulet 2, Hulet en Stoof, Schunck}. Today, the main
phases of this system, as a function of temperature, interaction strength and
polarization, have been observed experimentally \cite{Ketterle FD}. One of the
most fundamental effects that occurs in a polarized Fermi gas is that above a
critical imbalance, fermionic superfluidity ceases to exist and the system
makes a transition into a normal Fermi gas. This is known as the
Clogston-Chandrasekhar limit \cite{Clogston-Chandrasekhar}. One important
question that has emerged is whether there exist other pairing-mechanisms by
which a gas of interacting fermions can accommodate polarized superfluidity.
Recently, there has been an intensive theoretical search for these exotic
superfluid phases, such as the breached pair or Sarma phase \cite{Sarma, Sarma
Studies}, and phase separation \cite{Phase separation}. Another phase that has
attracted wide attention is the Fulde-Ferrell-Larkin-Ovchinnikov phase (FFLO
phase), that was proposed independently by Fulde and Ferrell \cite{Fulde
Ferell} and by Larkin and Ovchinnikov \cite{Larkin Ovchinnikov} in 1964. Their
idea was that a fermionic system can accommodate spin imbalance, by forming
pairs with finite center-of-mass momentum. Up till now, this state has never
been observed. A recent paper reports indirect experimental evidence for the
FFLO state in one dimension (1D) \cite{Liao Rittner}, but in three dimensions
(3D), the FFLO state has so far eluded experimental observation. This can be
related to theoretical predictions, which state that the FFLO state only
occupies a tiny sliver of the BCS-BEC crossover phase diagram \cite{Hu Liu,
Sheehy Radzihovsky}. It was therefore necessary to find a new method to
stabilize the FFLO state in 3D. At the moment, new ideas are emerging to
realize this goal. For instance, in two recent papers, it was suggested to
stabilize the FFLO state by the use of a 3D cubic lattice \cite{Koponen,
Trivedi}. As an alternative approach, we proposed to stabilize the FFLO state
in an imbalanced 3D Fermi gas by adding a 1D optical potential to this system
\cite{Devreese Klimin en Tempere (2011)}. This turned out to enhance the
stability region of the FFLO state by up to a factor six. In Ref.
\cite{Devreese Klimin en Tempere (2011)}, the proof of principle for this
concept was established using the path integral method , but questions
remained about the exact effects of the properties of the optical potential on
the FFLO state. In this paper, we investigate the role of the optical
potential in detail. Our aim here is to present results that relate closely to
the experiment. To this end, we construct and discuss the phase diagrams of an
imbalanced Fermi gas in 3D, subjected to a 1D optical potential, both at fixed
chemical potentials and at fixed densities. Subsequently, we investigate and
discuss the effects of the properties of the optical potential on the
BCS\ superfluid state and on the FFLO state. In Sec. \ref{Concept} we outline
our strategy to construct the phase diagrams at fixed chemical potentials and
at fixed densities. Subsequently, we discuss the competition between the BCS
phase and the FFLO phase in these phase diagrams, which differ qualitatively
from the corresponding phase diagrams for the imbalanced Fermi gas in 3D
without the optical potential. In Sec. \ref{Effects}, we investigate the
effects of the parameters of the optical potential on the FFLO state.
Furthermore, we explain the remarkable effect that the maximal polarization at
which the FFLO state can exist decreases when the interaction strength is
increased above a certain critical value. Finally in section \ref{Conclusion}
we draw conclusions.

\section{Construction of the phase diagrams\label{Concept}}

\subsection{Phase diagram at fixed chemical potentials}

To construct the phase diagram of an imbalanced 3D Fermi gas, subjected to a
1D optical potential, at fixed chemical potentials, we start from the
saddle-point free energy per unit volume of this system, which is given by
\cite{Devreese Klimin en Tempere (2011)}:%
\begin{align}
\Omega_{sp}\left(  \mu,\zeta;\Delta,Q\right)   &  =-\frac{1}{\left(
2\pi \right)  ^{2}}\int_{0}^{+\infty}dk~k~\int_{-Q_{L}}^{+Q_{L}}dk_{z}%
\nonumber \\
&  \times \left[  \max \left[  \left \vert \zeta_{k,Q}\right \vert ,E_{\mathbf{k}%
}\right]  -\xi_{\mathbf{k}}-\frac{\Delta^{2}}{2\left \{  k^{2}+\delta \left[
1-\cos \left(  \frac{\pi k_{z}}{Q_{L}}\right)  \right]  \right \}  }\right]
+\frac{\Delta^{2}}{8\pi} \label{free energy}%
\end{align}
with
\begin{equation}
\left \{
\begin{array}
[c]{l}%
\xi_{\mathbf{k}}=k^{2}+\delta \left[  1-\cos \left(  \frac{\pi}{2}\frac{Q}%
{Q_{L}}\right)  \cos \left(  \frac{\pi k_{z}}{Q_{L}}\right)  \right]  -\mu \\
E_{\mathbf{k}}=\sqrt{\xi_{\mathbf{k}}^{2}+\Delta^{2}}\\
\zeta_{k,Q}=\zeta-\delta \sin \left(  \frac{\pi}{2}\frac{Q}{Q_{L}}\right)
\sin \left(  \frac{\pi k_{z}}{Q_{L}}\right)
\end{array}
\right.  .
\end{equation}
To derive expression (\ref{free energy}), the 1D optical potential was
described by using a modified dispersion relation for the fermionic
particles:
\begin{equation}
\varepsilon(k,k_{z})=k^{2}+\delta \left[  1-\cos \left(  \frac{\pi k_{z}}{Q_{L}%
}\right)  \right]  , \label{dispersion}%
\end{equation}
where $\delta$ is a function of the depth of the potential $V_{0}$ and of the
recoil energy $E_{R}$, and $Q_{L}$ is the wave vector of the optical
potential. Expression (\ref{dispersion}) consists of a free particle
dispersion in the directions perpendicular to the optical potential, and a
tight-binding periodic dispersion \cite{Menotti} in the direction parallel to
the optical potential. Because we want to describe the FFLO state, we only
consider the BCS-side of the BCS-BEC crossover \cite{Iskin de Melo,
Radzihovsky}. In this region, the scattering length is always negative, hence
the following units are used (\ref{free energy}): $\hbar=2m=-a_{s}=1$, where
$a_{s}$ is the s-wave scattering length. Here it must be mentioned that the
derivation leading to the free energy (\ref{free energy}) neglected the
density modulation of the Fermi gas due to the 1D optical potential. This
approximation leads to an underestimation of the interactions between the
spin-up and spin-down fermions.

The free energy (\ref{free energy}) depends on two thermodynamic variables
$\mu$ and $\zeta$, and on two variational parameters $\Delta$ en $Q$. The
variables $\mu$ and $\zeta$ are the total chemical potential $\mu=\left(
\mu_{\uparrow}+\mu_{\downarrow}\right)  /2$, which fixes the total density,
and the imbalance chemical potential $\zeta=\left(  \mu_{\uparrow}%
-\mu_{\downarrow}\right)  /2$, which fixes the polarization. A description in
terms of $\mu$ and $\zeta$ is equivalent to a description in terms of the
respective chemical potentials$~$of both spin species $\mu_{\uparrow}$ and
$\mu_{\downarrow}$. The variational parameters $\Delta$ and $Q$ denote the
binding energy and the center-of-mass momentum of the fermion pairs
respectively. The parameter $Q$ was introduced to include the FFLO state
(which is characterized by fermion pairs with finite momentum) in our
description. To find the ground state of the system, given fixed values of
$\mu$ and $\zeta$, the saddle-point free energy $\Omega_{sp}\left(  \mu
,\zeta;\Delta,Q\right)  $ has to be minimized with respect to the variational
parameters $\Delta$ and $Q$. This defines the saddle-point equations%
\begin{equation}
\left \{
\begin{array}
[c]{c}%
\left.  \dfrac{\partial \Omega_{sp}\left(  \mu,\zeta;\Delta,Q\right)
}{\partial \Delta}\right \vert _{\mu,\zeta}=0\\
\left.  \dfrac{\partial \Omega_{sp}\left(  \mu,\zeta;\Delta,Q\right)
}{\partial Q}\right \vert _{\mu,\zeta}=0
\end{array}
\right.  , \label{gap equation}%
\end{equation}
which have to be satisfied simultaneously. One must be careful when
extremizing the free energy, because only the minima correspond to a stable
physical state.\ We avoided this difficulty by explicitly minimizing the free
energy, instead of searching for roots of (\ref{gap equation}). Depending on
the values of the parameters $\mu$ and $\zeta$, three different ground states
can emerge: 1) a spin-balanced (S-b) superfluid, with $\Delta \neq0$ and $Q=0$
, 2) the FFLO state with $\Delta \neq0$ and $Q\neq0$, and 3) the normal state
with $\Delta=0$. Moreover, in the phase diagrams we will make a distinction
between the partially polarized normal state (PPN) and the fully polarized
normal state (FPN). In the latter, all particles are in the same spin-state,
while in the former both spin-states are still present. Following this
procedure, the phase diagram as a function of $\mu$ and $\zeta$ can be
constructed. Before doing so, the interval of the chemical potential $\mu$
that is relevant to describe the FFLO state must be estimated. As mentioned
earlier, we only consider the BCS-side of the BCS-BEC crossover. This means
that the variable $\left(  k_{F}a_{s}\right)  ^{-1}$, with $k_{F}$ the Fermi
wave vector, which characterizes the interaction strength of a two-component
Fermi gas in 3D, has to lie in the interval $\left]  -\infty,0\right]  $. At
unitarity, $\left(  k_{F}a_{s}\right)  ^{-1}=0$, which means that, in our
description, the density has to go to infinity since we work in units where
$a_{s}=-1$. Even at $\left(  k_{f}a_{s}\right)  ^{-1}=0$, the chemical
potential is still a fraction of the Fermi energy, and since the Fermi energy
goes to infinity it follows that $\mu=+\infty$. In practice, it turns out that
$\mu \approx6$ lies close enough to unitarity to describe the full stability
region of the FFLO phase. The BCS-limit is already attained within good
approximation at $\left(  k_{F}a_{s}\right)  ^{-1}=-2$. In this limit,
$\mu=E_{F}$ from which it follows that $\mu=1/4$.

\begin{figure}
[h]
\begin{center}
\includegraphics[
trim=-0.014001in 0.139311in 0.014001in -0.139311in,
height=12.2967cm,
width=8.9029cm
]%
{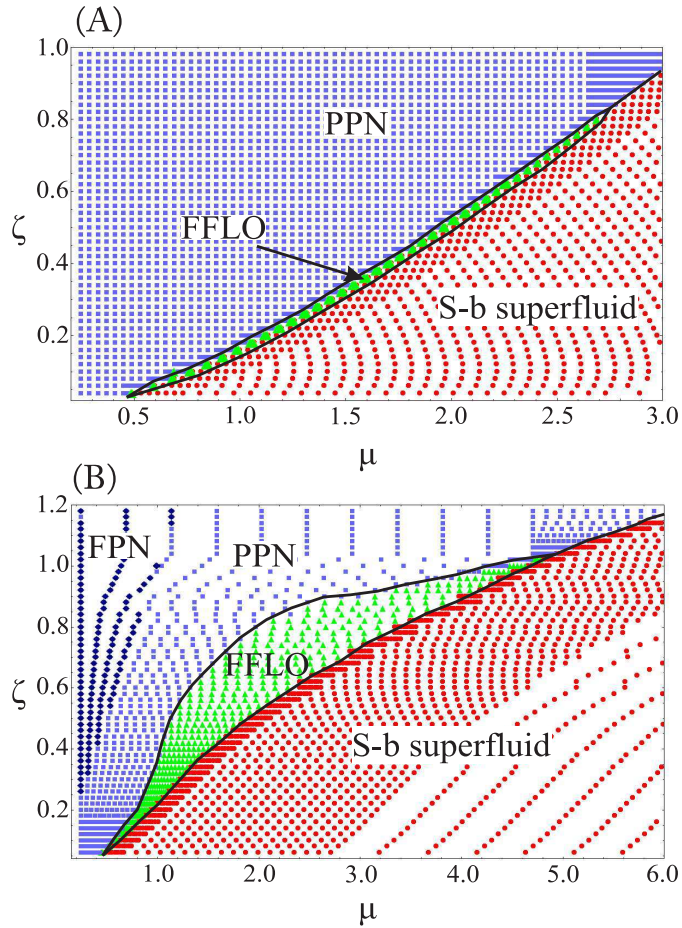}%
\caption{Comparison between the phase diagram in the $\left(  \mu
,\zeta \right)  $-plane ($\mu$ and $\zeta$ are in units $E_{0}=\frac{\hbar^{2}%
}{2ma_{s}^{2}}$) of (A) an imbalanced Fermi gas in 3D, and (B) an imbalanced
Fermi gas in 3D, subjected to a 1D optical potential. Both phase diagrams are
constructed based on a finite set of points in the $\left(  \mu,\zeta \right)
$-plane. For each point, the ground state of the system is determined. There
are four possible ground states: a spin-balanced (S-b) superfluid (red
circles), the FFLO phase (green triangles), the partially polarized normal
(PPN) phase (light blue squares) and the fully polarized normal (FPN) phase
(dark blue diamonds). The presence of the 1D optical potential leads to a
significant enlargement of the FFLO region. We used $s=\frac{V_{0}}{E_{R}}=5$
and $\lambda=1200$nm for the depth and wave length of the optical potential,
and $\left \vert a_{s}\right \vert =500$nm for the s-wave scattering length.}%
\label{figure1}%
\end{center}
\end{figure}

Figure \ref{figure1} shows a comparison between the phase diagrams in the
$\left(  \mu,\zeta \right)  $-plane of an imbalanced Fermi gas in 3D (figure
\ref{figure1} (A)), and an imbalanced Fermi gas in 3D, subjected to a 1D
optical potential (figure \ref{figure1} (B)).%
%TCIMACRO{\FRAME{fhFU}{8.9029cm}{12.2967cm}{0pt}{\Qcb{Comparison between the
%phase diagram in the $\left(  \mu,\zeta\right)  $-plane ($\mu$ and $\zeta$ are
%in units $E_{0}=\frac{\hbar^{2}}{2ma_{s}^{2}}$) of (A) an imbalanced Fermi gas
%in 3D, and (B) an imbalanced Fermi gas in 3D, subjected to a 1D optical
%potential. Both phase diagrams are constructed based on a finite set of points
%in the $\left(  \mu,\zeta\right)  $-plane. For each point, the ground state of
%the system is determined. There are four possible ground states: a
%spin-balanced (S-b) superfluid (red circles), the FFLO phase (green
%triangles), the partially polarized normal (PPN) phase (light blue squares)
%and the fully polarized normal (FPN) phase (dark blue diamonds). The presence
%of the 1D optical potential leads to a significant enlargement of the FFLO
%region. We used $s=\frac{V_{0}}{E_{R}}=5$ and $\lambda=1200$nm for the depth
%and wave length of the optical potential, and $\left\vert a_{s}\right\vert
%=500$nm for the s-wave scattering length.}}{\Qlb{figure1}}{figure1.eps}%
%{\special{ language "Scientific Word";  type "GRAPHIC";  display "USEDEF";
%valid_file "F";  width 8.9029cm;  height 12.2967cm;  depth 0pt;
%original-width 6.7966in;  original-height 10.1001in;  cropleft "-0.00206";
%croptop "1.013793";  cropright "0.99794";  cropbottom "0.013793";
%filename 'Figure1.eps';file-properties "XNPEU";}}}%
%BeginExpansion
%EndExpansion
In both these phase diagrams the ground state of the system is determined for
a discrete set of $\left(  \mu,\zeta \right)  $-points and each of these points
is displayed. We chose this type of visualization so that later on in this
paper, the transformation to the phase diagram at fixed densities will be
clarified by the transformation of the set of $\left(  \mu,\zeta \right)
$-points to a set of $\left(  \frac{1}{k_{F}a_{s}},P\right)  $-points, where
$P$ denotes the polarization. In both phase diagrams in figure \ref{figure1},
the system is always in a spin-balanced superfluid state when $\zeta=0$. This
is not visible in Fig. \ref{figure1} since in this figure $\zeta \geq0.06$. For
small values of the total chemical potential $\mu$ ($\mu \leq0.5$ in Figs.
\ref{figure1} (A) and (B)), the system makes a transition into a partially
polarized normal state when $\zeta$ becomes of the order of the gap $\Delta$.
For larger values of $\mu$, the system goes through all three different states
when the imbalance chemical potential $\zeta$ is increased from zero. For low
values of $\zeta$, the system remains a spin-balanced superfluid. This means
that a finite energy is needed in order to convert particles from one spin
species to the other. The minimal amount of energy that is required for this
conversion increases when the total chemical potential increases. For high
enough values of $\zeta$ the Fermi gas becomes polarized. At this point the
system makes a transition into the FFLO state, both in figure \ref{figure1}
(A) and (B). Characteristic of the FFLO state is that it has an oscillating
order parameter \cite{Fulde Ferell, Larkin Ovchinnikov}. This order parameter
is enhanced by the presence of the 1D potential, which results in a much
larger FFLO region in Fig. \ref{figure1} (B) when compared to the case without
the optical potential in Fig. \ref{figure1} (A). When the imbalance chemical
potential $\zeta$ increases further, the system will go over into a partially
polarized (PPN) state and eventually into a fully polarized (FPN) state, and
superfluidity is destroyed.

In the phase diagram in Fig. \ref{figure1}, we made the following choice for
the depth $s=V_{0}/E_{R}$ and for the wave length $\lambda$ of the optical
potential: $s=5$ and $\lambda=1000nm$, which are both realizable in current
experiments. The depth of the optical potential must be chosen carefully,
because the dispersion relation (\ref{dispersion}) that is used in our
description is only valid in the tight-binding limit. This will be discussed
in more detail in Sec. \ref{role of potential parameters}. Both parameters $s$
and $\lambda$ correspond to specific values of $\delta$ and $Q_{L}$, which are
defined in expression (\ref{dispersion}). Since we use units of scattering
length, a fixed value of this quantity has to be chosen, in order to compute
the values of $\delta$ and $Q_{L}$. Here we choose $\left \vert a_{s}%
\right \vert =500nm$, which corresponds to a situation close to a Feshbach resonance.

\subsection{Phase diagram at fixed densities}

\begin{figure}
[h]
\begin{center}
\includegraphics[
height=5.9748cm,
width=8.9029cm
]%
{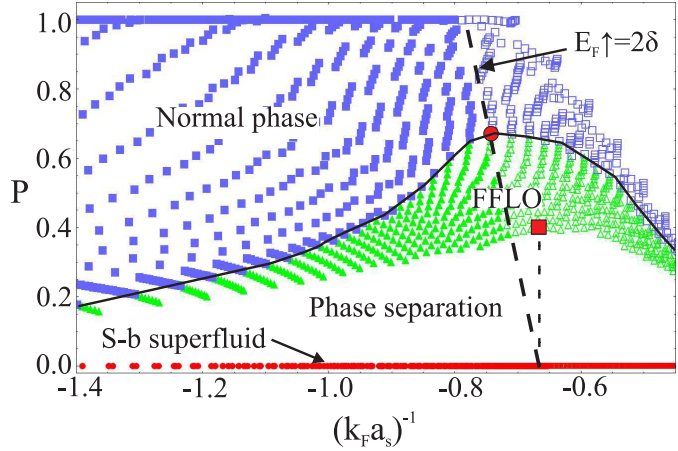}%
\caption{Phase diagram of an imbalanced Fermi gas in 3D, subjected to a 1D
optical potential, at fixed density and fixed polarization P. The
dimensionless parameter $\left(  k_{F}a_{s}\right)  ^{-1}$ characterizes the
interaction strength. Each point in this phase diagram is a transformation of
a corresponding point in the phase diagram in figure \ref{figure1} (B). At
polarization zero, the system is in the spin-balanced (S-b) superfluid phase.
When polarization becomes finite, there is a region where no points exist.
This means that no combination of density and polarization can solve the
number equations, and the system separates into two phases: a spin-balanced
superfluid component and an FFLO component. Interestingly, both the boundary
between the phase separation region and FFLO, and between FFLO and the normal
phase have a maximum (indicated by a red square and a red circle
respectively). The dashed oblique line indicates the boundary where the Fermi
energy of the majority spins becomes equal to the top of the first Bloch band,
which is equal to $2\delta$. The empty blue squares (normal phase) and empty
green triangles (FFLO) indicate that the Fermi energy of the majority spins
lies higher than the top of the first Bloch band. The vertical dashed line
indicates the value of the interaction strength where the Fermi energy at
polarization zero becomes equal to the top of the first Bloch band. Both
dashed lines intersect with one of the two indicated maxima. ($s=5,$
$\lambda=1200$nm, and $\left \vert a_{s}\right \vert =500$nm)}%
\label{figure2}%
\end{center}
\end{figure}

In the previous section, we chose to construct a phase diagram at fixed
chemical potentials $\mu$ and $\zeta$. This is not the only possible choice.
Another possibility is to work at fixed total density $n=n_{\uparrow
}+n_{\downarrow}$ and fixed polarization $P=\left(  n_{\uparrow}%
-n_{\downarrow}\right)  /\left(  n_{\uparrow}+n_{\downarrow}\right)  $, which
is equivalent to fixing the density of both spin species $n_{\uparrow}$ and
$n_{\downarrow}$. To transform between these two descriptions, the following
number equations have to be solved:%
\begin{equation}
\left \{
\begin{array}
[c]{c}%
n=-\left.  \dfrac{\partial \Omega_{sp}\left(  \mu,\zeta;\Delta,Q\right)
}{\partial \mu}\right \vert _{\zeta}\\
\delta n=-\left.  \dfrac{\partial \Omega_{sp}\left(  \mu,\zeta;\Delta,Q\right)
}{\partial \zeta}\right \vert _{\mu}%
\end{array}
\right.  . \label{number eqs.}%
\end{equation}
In principle, we need to use the chain rule in (\ref{number eqs.}), since
$\Delta$ and $Q$ are separate variables that depend on $\mu$ and $\zeta$. But
the saddle-point condition (\ref{gap equation}) will make these terms
disappear. If fluctuations around the saddle point are taken into account,
these extra terms will remain. The inclusion of fluctuations around the saddle
point lies beyond the scope of this paper. Starting from the phase diagram at
fixed chemical potentials in figure \ref{figure1} (B), the phase diagram at
fixed densities can now be constructed. Every point in the phase diagram
depicted in figure \ref{figure1} (B) corresponds to a set of values for the
parameters $\left(  \mu,\zeta;\Delta,Q\right)  $. Inserting these values into
both equations in (\ref{number eqs.}) yields the density $n$ and the density
difference $\delta n=n_{\uparrow}-n_{\downarrow}$. The latter can also be
written as the polarization $P=\delta n/n$. The final step is to find the
interaction strength $\left(  k_{F}a_{s}\right)  ^{-1}$ that corresponds to
the density $n$. Since the scattering length $a_{s}$ is kept constant, this
can be done by using the following expression that relates the density to the
Fermi wave vector $k_{F}$:%
\begin{equation}
n=\left \{
\begin{array}
[c]{c}%
\dfrac{Q_{L}}{2\pi^{2}}\left(  k_{F}^{2}-\delta \right)  ~\left(  E_{F}%
\geq2\delta \right) \\
\dfrac{Q_{L}}{2\pi^{3}}\left[  \left(  k_{F}^{2}-\delta \right)  \arccos \left(
\dfrac{\delta-k_{F}^{2}}{\delta}\right)  +\delta \sqrt{1-\left(  \dfrac
{\delta-k_{F}^{2}}{\delta}\right)  ^{2}}\right]  ~\left(  E_{F}<2\delta
\right)
\end{array}
\right.  . \label{density}%
\end{equation}
This expression (\ref{density}) is derived by calculating the number of
particles at temperature zero that have an energy that lies lower than the
Fermi energy, where the particles are assumed to have a dispersion relation
given by (\ref{dispersion}). The procedure outlined above yields the phase
diagram at fixed density and at fixed polarization, which is shown in figure
\ref{figure2}. Each point on this phase diagram corresponds to a point on the
$\left(  \mu,\zeta \right)  $-phase diagram in figure \ref{figure1} (B).%
%EndExpansion
In the case of zero polarization, the system is in the spin-balanced
superfluid state. Figure \ref{figure2} shows that there exists a portion of
the phase diagram at finite polarization, where no points are present.
Mathematically, this corresponds to the case when there exists no combination
of density and polarization that can satisfy both number equations
(\ref{number eqs.}). Physically, this is a state where the system separates
into two different phases: a spin-balanced superfluid phase and an FFLO phase,
where the latter is being pushed out of the former. When polarization
increases further, the system makes a transition into the FFLO state. Figure
\ref{figure2} also shows that when the interaction strength increases, both
the boundary between the phase separation region and the FFLO state and the
boundary between the FFLO state and the normal state, lie at higher values of
polarization. Interestingly though, for both boundaries there exists a
critical interaction strength at which this polarization becomes maximal.
These two maxima are indicated by a red square and a red circle respectively
in figure \ref{figure2}. If the interaction strength becomes larger than these
two critical values, both the region of phase separation and the FFLO region
seem to shrink. This counterintuitive effect will be discussed in the next
section. When polarization increases further, the energy cost of forming pairs
with finite momentum will eventually become too high and the system will make
a transition into a normal Fermi gas. The points with polarization $P=1$
correspond to the fully polarized region in Fig. \ref{figure1}. We now have
given a general overview of the phase diagrams of an imbalanced 3D Fermi gas
subjected to a 1D potential. In the next section we will investigate the
effects of the properties of the optical potential on the FFLO state.

\section{Effects of the 1D potential on the FFLO state \label{Effects}}

\subsection{Role of the wave length and the depth of the optical
potential\label{role of potential parameters}}

There are two parameters that determine the properties of the 1D optical
potential: the potential depth $s$, and the wave length $\lambda$. Before
discussing the effect of these parameters, it is important to know to which
intervals they are limited in our current treatment. For the wave length, we
take realistic values corresponding to present day experiments: $\lambda
\in \left[  600nm,1200nm\right]  $. We have to be careful, however, when
choosing values for the depth of the optical potential. There are two physical
reasons to set a lower boundary to this depth. Firstly, since the optical
potential is described by a tight-binding dispersion relation (Eq.
\ref{dispersion}), the depth of the potential must be large enough to justify
this approximation. To quantify this, we calculated the energy spectrum of a
particle in a periodic potential, and compared it to the tight-binding
dispersion given by (\ref{dispersion}). For completeness, we added a concise
overview of this calculation in appendix \ref{appendix calc spectrum}. This
calculation shows that the error made by using expression (\ref{dispersion})
for the dispersion is less than five percent when $s\geq4$. Secondly, by using
the dispersion relation (\ref{dispersion}), we only take into account the
lowest Bloch band. This description is justified as long as the system has a
Fermi energy which lies below the second Bloch band. A Fermi energy that lies
between the top of the first Bloch band and the bottom of the second Bloch
band is allowed, since the dispersion in the direction perpendicular to the
direction of the optical potential is still the same as in the free 3D Fermi
gas case. We have verified that by considering a potential depth $s\geq4$, the
system has a Fermi energy that lies below the bottom of the second Bloch band,
for all cases considered in this paper. Based on this analysis, we will use
the lower boundary $s=4$ for the potential depth throughout the rest of this
paper. There is also a reason to set an upper limit to the depth of the
optical potential. When the potential becomes too deep, the 3D Fermi gas will
transition into a series of weakly-coupled two-dimensional (2D) pancakes. Such
a 2D Fermi gas must be described using a different renormalized interaction
strength \cite{Artikels 2D tunneling}. We therefore choose $s$ to lie between
4~and 6, so that both the 3D description and the tight-binding approximation
are justified.

\begin{figure}
[h]
\begin{center}
\includegraphics[
height=9.1336cm,
width=13.3203cm
]%
{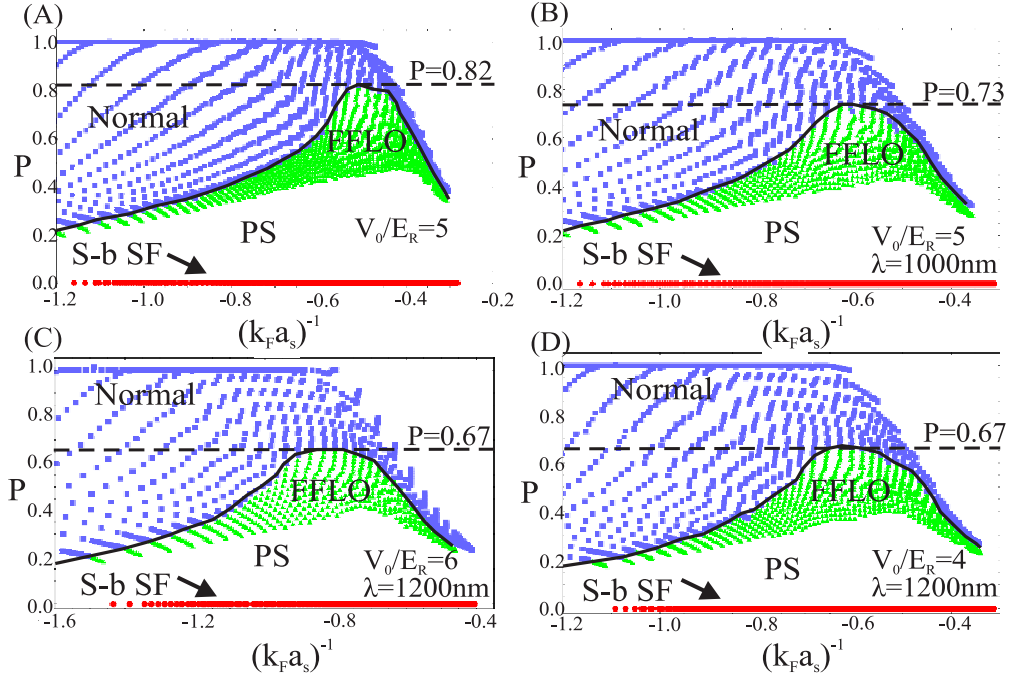}%
\caption{Overview of several phase diagrams of an imbalanced Fermi gas in 3D
subjected to a 1D optical potential, for different values of the depth $s$ and
the wave length $\lambda$ of the potential. The maximal polarization P at
which the FFLO state can exist (indicated by a dashed horizontal line)
increases when the wavelength of the optical potential becomes smaller. At
constant wave length, the depth of the optical potential does not have an
influence on this maximal polarization, at least within the range $4\leq
s\leq6$ that is considered here. (S-b SF = spin-balanced superfluid, PS =
phase separation)}%
\label{figure3}%
\end{center}
\end{figure}

We now proceed to investigate the effects of both optical potential parameters
$\lambda$ and $s$. Figure \ref{figure3} shows several phase diagrams at fixed
density and polarization, for an imbalanced Fermi gas in 3D subjected to a 1D
potential, for various values of both parameters $\lambda$ and $s$.
%TCIMACRO{\FRAME{fhFU}{13.3203cm}{9.1336cm}{0pt}{\Qcb{Overview of several phase
%diagrams of an imbalanced Fermi gas in 3D subjected to a 1D optical potential,
%for different values of the depth $s$ and the wave length $\lambda$ of the
%potential. The maximal polarization P at which the FFLO state can exist
%(indicated by a dashed horizontal line) increases when the wavelength of the
%optical potential becomes smaller. At constant wave length, the depth of the
%optical potential does not have an influence on this maximal polarization, at
%least within the range $4\leq s\leq6$ that is considered here. (S-b SF =
%spin-balanced superfluid, PS = phase separation)}}{\Qlb{figure3}}%
%{figure3.eps}{\special{ language "Scientific Word";  type "GRAPHIC";
%width 13.3203cm;  height 9.1336cm;  depth 0pt;  original-width 7.3215in;
%original-height 5.0081in;  cropleft "0";  croptop "1";  cropright "1";
%cropbottom "0";  filename 'Figure3.eps';file-properties "XNPEU";}}}%
%BeginExpansion
%EndExpansion
In Figs. \ref{figure3} (A), (B), and (C) the wave length of the optical
potential is varied. These three figures show that the maximal polarization at
which the FFLO state can exist increases when the wave length of the optical
potential becomes smaller. This effect agrees with our expectations and can
best be explained in terms of the Fermi surfaces in momentum space. When an
imbalance exists between spin-up and spin-down particles, an energy gap opens
up between the Fermi surfaces of the two spin species. This is akin to a
Zeeman-splitting effect. In order to bridge this gap and to let particles pair
with the same energy, the FFLO state forms pairs with a finite momentum $Q$.
The larger the imbalance, the larger the energy gap becomes and the larger $Q$
has to be. In Ref. \cite{Devreese Klimin en Tempere (2011)} we demonstrated
that the wave vector of the FFLO state cannot become larger than the wave
vector of the optical potential. Since a smaller wave length corresponds to a
larger wave vector, the maximal value of the wave vector $Q$ that can be
attained by the FFLO state increases when the wave length of the optical
potential decreases. This means that a larger energy gap can be bridged and
thus a larger polarization can be accommodated. Figures \ref{figure3} (C) and
\ref{figure3} (D) show the phase diagram of the system at the same wave length
but at different potential depth. In both figures, the maximal polarization of
the FFLO state is equal. We conclude that only the wave length of the optical
potential has an influence on the capability of the system to remain in the
FFLO state while being polarized, at least within the small range of potential
depth $s$ that is considered here.

\subsection{Effect of the edge of the Brillouin zone}

One effect that is visible throughout all phase diagrams in figures
\ref{figure2} and \ref{figure3}, is that when the interaction strength exceeds
a certain critical value, both the region of phase separation and the FFLO
region become less stable with respect to polarization (i.e. the maximal
polarization at which they can exist, decreases) . This can be seen by the
fact that both the boundary between the phase separation region and the FFLO
state and the boundary between the FFLO state and the normal state have a
maximum (indicated by a red square and a red circle respectively in figure
\ref{figure2}). Moreover, these two maxima seem to correspond to different
values of the interaction strength. This effect is counterintuitive, because
when the interaction strength increases, the binding energy of the fermion
pairs increases likewise, which in general should strengthen superfluidity,
not weaken it.

\begin{figure}
[h]
\begin{center}
\includegraphics[
height=9.0808cm,
width=8.9029cm
]%
{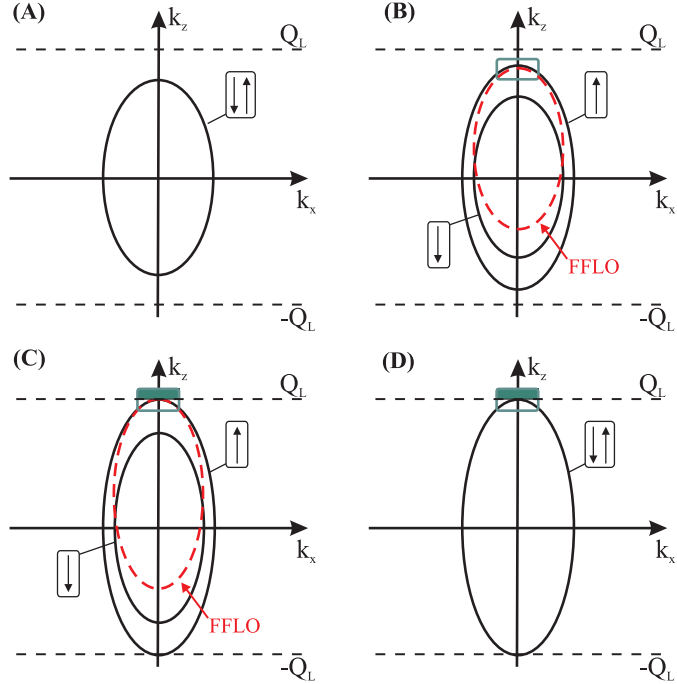}%
\caption{Schematic representation of the pairing mechanisms in an imbalanced
3D Fermi gas subjected to a 1D optical potential. For the sake of clarity, we
have omitted the y direction. (A) The Fermi surfaces of the spin-up and the
spin-down particles at polarization zero. Pairing occurs in a band with
thickness $\Delta$ around the Fermi energy and BCS superfluidity is possible.
(B) When polarization is introduced, a gap is created between the Fermi
surfaces of both spin species. If this gap becomes too big, there will be no
overlap between the pairings bands around the Fermi surfaces. The system can
solve this by translating the minority Fermi surface (dashed red ellipsoid) so
that locally, both Fermi surfaces are re-aligned (indicated by a green
rectangle). (C) When the density increases, the Fermi surface of the majority
spins (here spin-up) will lie at the edge of the first Brillouin zone. This
will hinder FFLO-like pairing because no states are available at $k_{z}>Q_{L}%
$. (D) Above a critical density, the Fermi surface lies at the edge of the
Brillouin zone even for polarization zero. At this point, BCS superfluidity
will also be hindered.}%
\label{figure4}%
\end{center}
\end{figure}

This effect is explained with the help of figure \ref{figure2}. In this
figure, the empty blue squares (normal state) and the empty green triangles
(FFLO state) indicate that the Fermi energy of the majority spins lies higher
than the top of the first Bloch band, which is equal to $2\delta$. The
boundary at which the Fermi energy of the majority spins becomes equal to the
top of the first Bloch band is indicated by the dashed oblique line. The
maximum polarization under which the FFLO state can exist (red circle), lies
exactly at this line. Furthermore, the vertical dashed line indicates the
value of the interaction strength at which the Fermi energy at polarization
zero becomes equal to the top of the first Bloch band. This dashed vertical
line intersects with the maximum polarization under which the phase separation
region can exist (red square). The reason why each of these two dashed lines
intersects with one of the two maxima can be explained as follows. Consider
the case which is shown schematically in Fig. \ref{figure4}. Figure
\ref{figure4} (A) shows the Fermi surface of both spin-up and spin-down
particles at polarization zero. In order for the system to become superfluid,
pairing occurs between particles that lie in a band with thickness $\Delta$
around the Fermi energy. When these pairing bands of the spin-up and spin-down
particles overlap, pairing is possible and the system is in the superfluid
state. When polarization is introduced, a gap arises between the two Fermi
surfaces, as indicated in figure \ref{figure4} (B). As long as this gap is
smaller than $\Delta$, there is overlap of the pairing bands, and
superfluidity is possible in principle. When the polarization becomes too
large, the overlap disappears and BCS superfluidity is no longer possible. In
order to remain superfluid, the system can transition into the FFLO state,
which translates the Fermi surface of the minority spins so that it locally
lies on the Fermi surface of the majority spins, making pairing possible again
(see Fig. \ref{figure4} (B)). This allows the system to remain superfluid, at
the cost of paying some energy to form particles with finite center-of-mass
momentum. This is the case for a 3D Fermi gas, but when a 1D potential is
added, an additional effect comes into play. When the density and/or
polarization is high enough so that the Fermi surface of the majority spins
lies at the edge of the first Brillouin zone (this is true for the area at the
right side of the dashed oblique line in Fig. \ref{figure2}), pairing at this
Fermi surface becomes more difficult, because for $k_{z}>Q_{L}$, there are no
available states (see Fig. \ref{figure4} (C)). Hence, the formation of an
FFLO-type superfluid phase will be hindered. This now explains why, at equal
polarization, the normal phase becomes favored over the FFLO phase when the
interaction strength exceeds the value corresponding to the situation where
the Fermi energy of the majority spins equals the top of the first Bloch band
(indicated by a red circle in Fig. \ref{figure2}). The question can be raised
whether the FFLO state can still exist if the fermion pairs acquire a momentum
in the $x$ or $y$ direction, since there exists no Brillouin zone in these
directions. We did not treat this case in our paper because we assumed the
FFLO momentum to be of the form $Q=\left(  0,0,Q_{z}\right)  $.

In the case of figure \ref{figure4} (C), BCS superfluidity can still exist,
because when the Fermi surface of the minority spins is not translated,
pairing does not occur at the boundary of the first Brillouin zone. However,
when the density becomes high enough, even the Fermi surface of the particles
at zero polarization will lie at this edge. This situation is depicted in
figure \ref{figure4} (D). For this and for higher densities, pairing of BCS
superfluidity will also occur at the edge of the Brillouin zone and it will
suffer the same problem as the FFLO state. This effect can be seen in Fig.
\ref{figure2} from the fact that, at equal polarization, the FFLO state
becomes favored over the spin-balanced superfluid phase, when the value of the
interaction strength increases above the value corresponding to the situation
where the Fermi energy at zero polarization equals the top of the first Bloch
band (indicated by a red square in Fig. \ref{figure2}).%

%TCIMACRO{\FRAME{fhFU}{8.9029cm}{9.0808cm}{0pt}{\Qcb{Schematic representation
%of the pairing mechanisms in an imbalanced 3D Fermi gas subjected to a 1D
%optical potential. For the sake of clarity, we have omitted the y direction.
%(A) The Fermi surfaces of the spin-up and the spin-down particles at
%polarization zero. Pairing occurs in a band with thickness $\Delta$ around the
%Fermi energy and BCS superfluidity is possible. (B) When polarization is
%introduced, a gap is created between the Fermi surfaces of both spin species.
%If this gap becomes too big, there will be no overlap between the pairings
%bands around the Fermi surfaces. The system can solve this by translating the
%minority Fermi surface (dashed red ellipsoid) so that locally, both Fermi
%surfaces are re-aligned (indicated by a green rectangle). (C) When the density
%increases, the Fermi surface of the majority spins (here spin-up) will lie at
%the edge of the first Brillouin zone. This will hinder FFLO-like pairing
%because no states are available at $k_{z}>Q_{L}$. (D) Above a critical
%density, the Fermi surface lies at the edge of the Brillouin zone even for
%polarization zero. At this point, BCS superfluidity will also be hindered.}%
%}{\Qlb{figure4}}{figure4.eps}{\special{ language "Scientific Word";
%type "GRAPHIC";  maintain-aspect-ratio TRUE;  display "USEDEF";
%valid_file "F";  width 8.9029cm;  height 9.0808cm;  depth 0pt;
%original-width 7.0828in;  original-height 6.9003in;  cropleft "0";
%filename 'Figure4.eps';file-properties "XNPEU";}}}%
%BeginExpansion
%EndExpansion

\section{Conclusions \label{Conclusion}}

We have investigated the effects of a 1D optical potential on the FFLO state
in an imbalanced Fermi gas in 3D. To allow for a direct connection with the
experiments, we constructed the phase diagram of the system, both at fixed
chemical potentials and at fixed densities of both spin species. Subsequently,
we investigated the effects of the depth and the wave length of the optical
potential on the FFLO state. It was shown that this state can exist under a
larger polarization, when the wave length of the optical potential becomes
smaller. The depth of the optical potential did not have any appreciable
effect, within the range that can be considered in our tight-binding
approximation. By studying the phase diagram at fixed density we have
discovered an unexpected effect. When the interaction strength exceeds a
certain critical value, the maximal polarization at which both the FFLO phase
and the phase separation region can exist decreases. The underlying reason was
discussed in detail. We have found that when the density of the system becomes
so high that the Fermi energy exceeds the top of the lowest Bloch band,
superfluid pairing becomes hindered. This is because in that case, particles
with the highest energy lie at the edge of the first Brillouin zone. Since no
states exist outside of this zone, pairing is not possible and the superfluid
state is suppressed. We expect this effect, along with the enhancement of the
FFLO state, to be observable in experiments where a 1D optical potential is
applied to an imbalanced Fermi gas in 3D.

\begin{acknowledgments}
Acknowledgments -- The authors would like to thank Gerard 't Hooft, Carlos
S\'{a} de Melo, and Sergei Klimin for fruitful discussions. JPAD wishes to
thank Nick Van den Broeck for helpful discussions. One of the authors (JPAD)
acknowledges a Ph. D. fellowship of the Research Foundation - Flanders (FWO).
This work was supported by FWO-V projects G.0356.06, G.0370.09N, G.0180.09N, G.0365.08.
\end{acknowledgments}

\appendix{}

\section{Range of optical potential depths that allow the tight-binding
approximation\label{appendix calc spectrum}}

\begin{figure}
[h]
\begin{center}
\includegraphics[
height=9.3356cm,
width=13.3203cm
]%
{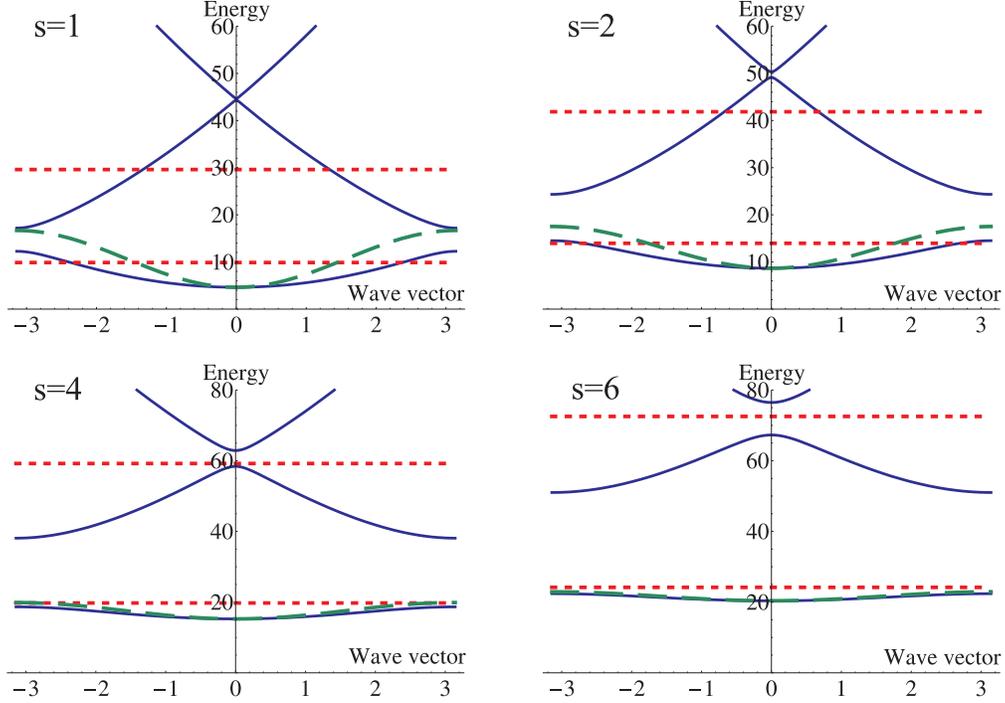}%
\caption{Comparison between the exact energy spectrum of a particle in a
periodic potential (solid blue lines) and the dispersion relation calculated
in the tight-binding limit (long-dashed green lines), given by expression
(\ref{dispersion}). The horizontal short-dashed red lines indicate the two
lowest energy levels of the harmonic oscillator, derived by approximating
(\ref{dispersion}) to second order in the momentum. For $s\geq4$, the
difference between the exact energy spectrum and the tight-binding dispersion
is less than 5\%.}%
\label{figure5}%
\end{center}
\end{figure}

The energy spectrum of a particle in a periodic potential can be calculated
using Schr\"{o}dinger's equation. The potential under consideration is given
by
\begin{equation}
V\left(  z\right)  =sE_{R}\sin^{2}\left(  Q_{L}z\right)  ,
\label{optische potentiaal}%
\end{equation}
where $s$ is the depth of the optical potential in units of the recoil energy
$E_{R}$ and $Q_{L}$ is the wave vector of the optical potential given by
$Q_{L}=\pi/\lambda$. Given this potential, Schr\"{o}dinger's equation states
that%
\begin{equation}
\left[  -\frac{\hbar^{2}}{2m}\frac{d^{2}}{dz^{2}}+sE_{R}\left(  \frac
{1-\cos \left(  2Q_{L}z\right)  }{2}\right)  \right]  \psi_{k}\left(  z\right)
=\varepsilon_{k}\psi_{k}\left(  z\right)  . \label{Schroedinger vergelijking}%
\end{equation}
Since we consider a particle in a periodic potential, Bloch's theorem applies
and we can re-write the wave function $\psi_{k}\left(  z\right)  $ as%
\begin{equation}
\psi_{k}\left(  z\right)  =e^{ikz}u_{k}\left(  z\right)  ,
\end{equation}
where $u_{k}\left(  z\right)  $ has the same periodicity as the potential.
Because $u_{k}\left(  z\right)  $ is periodic, it can be written in a Fourier
series%
\begin{align}
\psi_{k}\left(  z\right)   &  =e^{ikz}\left(  \sum_{n}c_{n}e^{i\frac{2\pi
n}{\lambda}z}\right) \nonumber \\
&  =\sum_{n}c_{n}e^{i\left(  k+2nQ_{L}\right)  z}.
\label{Bloch functie in Fourier}%
\end{align}
Substituting (\ref{Bloch functie in Fourier}) in Eq.
(\ref{Schroedinger vergelijking}) yields%
\begin{align}
&  \sum_{n}\left[  \frac{\hbar^{2}}{2m}\left(  k+2nQ_{L}\right)  ^{2}%
+\frac{sE_{R}}{2}\right]  c_{n}e^{i\left(  k+2nQ_{L}\right)  z}-\frac{sE_{R}%
}{2}\sum_{n}\left(  \frac{e^{2iQ_{L}z}+e^{-2iQ_{L}z}}{2}\right)
c_{n}e^{i\left(  k+2nQ_{L}\right)  z}\nonumber \\
&  =\varepsilon_{k}\sum_{n}c_{n}e^{i\left(  k+2nQ_{L}\right)  z}.
\label{vergelijking 1}%
\end{align}
Re-indexing the second sum over $n$ and identifying the coefficients on both
sides in equation (\ref{vergelijking 1}) leads to%
\begin{equation}
\left[  \frac{\hbar^{2}}{2m}\left(  k+2nQ_{L}\right)  ^{2}+\frac{sE_{R}}%
{2}-\varepsilon_{k}\right]  c_{n}-\frac{sE_{R}}{4}\left(  c_{n+1}%
+c_{n-1}\right)  =0. \label{centrale vergelijking}%
\end{equation}
This expression represents an infinite set of linear equations. In practice,
it is enough to consider five equations, since the contributions of the
eigenvectors $c_{n}$ decreases exponentially for large $n$. Solving equation
(\ref{centrale vergelijking}) numerically yields the exact energy spectrum of
a particle in a periodic potential, given by (\ref{optische potentiaal}).
Figure \ref{figure5} shows a comparison between this exact spectrum and the
corresponding dispersion relation in the tight-binding limit, given by
(\ref{dispersion}). For $s\leq2$ the difference is substantial, but for
$s\geq4$, the approximation deviates less than 5 percent from the exact
result. This justifies the approximation that we make by considering $s\geq
4$.
%TCIMACRO{\FRAME{fhFU}{13.3203cm}{9.3356cm}{0pt}{\Qcb{Comparison between the
%exact energy spectrum of a particle in a periodic potential (solid blue lines)
%and the dispersion relation calculated in the tight-binding limit (long-dashed
%green lines), given by expression (\ref{dispersion}). The horizontal
%short-dashed red lines indicate the two lowest energy levels of the harmonic
%oscillator, derived by approximating (\ref{dispersion}) to second order in the
%momentum. For $s\geq4$, the difference between the exact energy spectrum and
%the tight-binding dispersion is less than 5\%.}}{\Qlb{figure5}}{figure5.eps}%
%width 13.3203cm;  height 9.3356cm;  depth 0pt;  original-width 5.7917in;
%original-height 4.0499in;  cropleft "0";  croptop "1";  cropright "1";
%cropbottom "0";  filename 'Figure5.eps';file-properties "XNPEU";}}}%
%BeginExpansion
%EndExpansion


\begin{thebibliography}{99}                                                                                               %


\bibitem {Review Dalibard}I. Bloch, J. Dalibard, and W. Zwerger, Rev. Mod.
Phys. \textbf{80}, 885 (2008).

\bibitem {Greiner}M. Greiner, C. A. Regal, and D. S. Jin, Nature (London)
\textbf{426}, 537 (2003).

\bibitem {Chin}C. Chin, M. Bartenstein, A. Altmeyer, S. Riedl, S. Jochim, J.
Hecker Denschlag, R. Grimm, Science \textbf{305}, 1128 (2004).

\bibitem {Zwierlein Ketterle}M. W. Zwierlein, C. A. Stan, C. H. Schunck, S. M.
F. Raupach, A. J. Kerman, and W. Ketterle, Phys. Rev. Lett. \textbf{92,
}120403 (2004).

\bibitem {Dalibard Salomon}T.-L. Dao, A. Georges, J. Dalibard, C. Salomon, and
I. Carusotto, Phys. Rev. Lett. \textbf{98}, 240402 (2007).

\bibitem {Bloch review}I. Bloch, Nat. Phys. \textbf{1}, 23 (2005).

\bibitem {Greiner Q and A}M. Greiner and S. F\"{o}lling, Nature (London)
\textbf{453}, 736 (2008).

\bibitem {Chin Grimm}C. Chin, R. Grimm, P. Julienne, and E. Tiesinga, Rev.
Mod. Phys. \textbf{82}, 1225 (2010).

\bibitem {Regal Greiner}C. A. Regal, M. Greiner, and D. S. Jin, Phys. Rev.
Lett. \textbf{92}, 040403 (2004).

\bibitem {Bourdel Salomon}T. Bourdel, L. Khaykovich, J. Cubizolles, J. Zhang,
F. Chevy, M. Teichmann, L. Tarruell, S. J. J. M. F. Kokkelmans, and C.
Salomon, Phys. Rev. Lett. \textbf{93}, 050401 (2004).

\bibitem {Partridge Hulet}G. B. Partridge, K. E. Strecker, R. I. Kamar, M. W.
Jack, and R. G. Hulet, Phys. Rev. Lett. \textbf{95}, 020404 (2005).

\bibitem {De Melo}C. A. R. S\'{a} de Melo, M. Randeria and, J. R. Engelbrecht,
Phys. Rev. Lett. \textbf{71}, 3202 (1993).

\bibitem {Zwierlein Ketterle 2}M. W. Zwierlein, A. Schirotzek, C. H. Schunck,
and W. Ketterle, Science \textbf{311}, 492 (2006); M. W. Zwierlein, C. H.
Schunck, A. Schirotzek, and W. Ketterle, Nature (London) \textbf{442}, 54 (2006).

\bibitem {Partridge Hulet 2}G. B. Partridge, W. Li, R. I. Kamar, Y. A. Liao,
and R. G. Hulet,\ Science \textbf{311}, 503 (2006).

\bibitem {Shin Ketterle}Y. Shin, M. W. Zwierlein, C. H. Schunck, A.
Schirotzek, and W. Ketterle,\textit{\ }Phys. Rev. Lett. \textbf{97}, 030401 (2006).

\bibitem {Hulet en Stoof}G. B. Partridge, W. Li, Y. A. Liao, R. G. Hulet, M.
Haque, and H. T. C. Stoof, Phys. Rev. Lett. \textbf{97}, 190407 (2006).

\bibitem {Schunck}C. H. Schunck, Y. Shin, A. Schirotzek, M. W. Zwierlein, and
W. Ketterle\textit{, }Science \textbf{316}, 867 (2007).

\bibitem {Clogston-Chandrasekhar}A. M. Clogston, Phys. Rev. Lett. \textbf{9},
266 (1962); B. S. Chandrasekhar, Appl. Phys. Lett. \textbf{1}, 7 (1962).

\bibitem {Ketterle FD}W. Ketterle, Y. Shin, A. Schirotzek, and C. H. Shunk, J.
Phys. Condens. Matter \textbf{21}, 164206 (2009).

\bibitem {Sarma}G. Sarma, J. Phys. Chem. Solids \textbf{24}, 1029 (1963) ; W.
V. Liu and F. Wilczek, Phys. Rev. Lett. \textbf{90}, 047002 (2003).

\bibitem {Sarma Studies}K. B. Gubbels, M. W. J. Romans, and H. T. C. Stoof,
Phys. Rev. Lett. \textbf{97, }210402 (2006); L. He and P. Zhuang, Phys. Rev.
B. \textbf{79}, 024511 (2009).

\bibitem {Phase separation}P. F. Bedaque, H.\ Caldas, and G. Rupak, Phys. Rev.
Lett. \textbf{91}, 247002 (2003); J. Tempere, S. N. Klimin, and J. T.
Devreese, Phys. Rev. A \textbf{78}, 023626 (2008).

\bibitem {Fulde Ferell}P. Fulde and R. A. Ferrell, Phys. Rev. \textbf{135},
A550 (1964).

\bibitem {Larkin Ovchinnikov}A. I. Larkin and Y. N. Ovchinnikov, Zh. Eksp.
Teor. Fiz. \textbf{47}, 1136 (1964) [Sov. Phys. JETP \textbf{20}, 762 (1965)].

\bibitem {Liao Rittner}Y. Liao, A. S. C. Rittner, T. Paprotta, W. Li, G. B.
Partridge, R. G. Hulet, S. K. Baur, and E. J. Mueller, Nature \textbf{467},
567 (2010).

\bibitem {Hu Liu}H. Hu and X. J. Liu,\textit{\ }Phys. Rev. A \textbf{73},
051603(R) (2006).

\bibitem {Sheehy Radzihovsky}D. E. Sheehy and L. Radzihovsky, Ann. of Phys.
\textbf{322}, 1790 (2007); L. Radzihovsky and D. E. Sheehy, Rep. Prog. Phys.
\textbf{73, }076501 (2010).

\bibitem {Koponen}T. K. Koponen, T. Paananen, J.-P. Martikainen, and P.
T\"{o}rm\"{a}, Phys. Rev. Lett. \textbf{99,} 120403 (2007).

\bibitem {Trivedi}Y. L. Loh and N. Trivedi, Phys. Rev. Lett. \textbf{104,
}165302 (2010).

\bibitem {Devreese Klimin en Tempere (2011)}J. P. A. Devreese, S. N. Klimin,
and J. Tempere, Phys. Rev. A \textbf{83}, 013606 (2011).

\bibitem {Kleinert boek}H. Kleinert, \textit{Path Integrals in Quantum
Mechanics, Statistics, Polymer Physics, and Financial Markets}, 5th ed.(World
Scientific, Singapore, 2002).

\bibitem {Menotti}L. Pezz\`{e}, L. Pitaevskii, A. Smerzi, S. Stringari, G.
Modugno, E. de Mirandes, F. Ferlaino, H. Ott, G. Roati, and M. Inguscio, Phys.
Rev. Lett. \textbf{93}, 120401 (2004).

\bibitem {Iskin de Melo}M. Iskin and C. A. R. S\'{a} de Melo, Phys. Rev. B
\textbf{97}, 100404 (2006).

\bibitem {Radzihovsky}D. E. Sheehy and L. Radzihovsky, Phys. Rev. Lett.
\textbf{96}, 060401 (2006).

\bibitem {Artikels 2D tunneling}M. Wouters, J. Tempere, and J. T. Devreese,
Phys. Rev. A \textbf{70}, 013616 (2004); J. Tempere, J. Low. Temp. Phys.
\textbf{150,} 636 (2008).
\end{thebibliography}
\end{document}